\documentclass[aps,twocolumn,pre,superscriptaddress]{revtex4-2}

\usepackage{xcolor,hyperref}
\hypersetup{
   colorlinks,
   linkcolor={blue!50!black},
   citecolor={blue!50!black},
   urlcolor={blue!80!black}
}

\usepackage{amsmath}
\usepackage{amssymb}
\usepackage{color}
\usepackage{graphicx}

\DeclareMathAlphabet{\mathitbf}{OML}{cmm}{b}{it}
\newcommand{\zerovector}{\mathBold 0}

\newcommand{\zv}{\mathitbf z}

\newcommand{\xv}{\mathitbf x}
\newcommand{\uv}{\mathitbf u}

\newcommand{\piv}{\mathBold\pi}
\newcommand{\psiv}{\mathBold\psi}
\newcommand{\zetav}{\mathBold\zeta}
\newcommand{\calBold}[1]{\mbox{\boldmath${\cal #1}$}}
\newcommand{\mathBold}[1]{\mbox{\boldmath$#1$}}

\newcommand{\tripleCdot}{:\!\cdot\,}

\begin{document}

\title{Simple and Broadly Applicable Definition of Shear Transformation Zones}

\author{David Richard}
\thanks{contributed equally}
\affiliation{Institute for Theoretical Physics, University of Amsterdam, Science Park 904, Amsterdam, Netherlands}
\affiliation{Department of Physics, Syracuse University, Syracuse, NY 13244}
\author{Geert Kapteijns}
\thanks{contributed equally}
\affiliation{Institute for Theoretical Physics, University of Amsterdam, Science Park 904, Amsterdam, Netherlands}
\author{Julia A.~Giannini}
\affiliation{Department of Physics, Syracuse University, Syracuse, New York 13244}
\author{M.~Lisa Manning}
\affiliation{Department of Physics, Syracuse University, Syracuse, New York 13244}
\author{Edan Lerner}
\email{e.lerner@uva.nl}
\affiliation{Institute for Theoretical Physics, University of Amsterdam, Science Park 904, Amsterdam, Netherlands}

\begin{abstract}
Plastic deformation in amorphous solids is known to be carried by stress-induced localized rearrangements of a few tens of particles, accompanied by the conversion of elastic energy to heat. Despite their central role in determining how glasses yield and break, the search for a simple and generally applicable definition of the \emph{precursors} of those plastic rearrangements --- the so-called shear transformation zones (STZs) --- is still ongoing. Here we present a simple definition of STZs --- based solely on the \emph{harmonic} approximation of a glass' energy. We explain why and demonstrate directly that our proposed definition of plasticity carriers in amorphous solids is more broadly applicable compared to anharmonic definitions put forward previously. Finally, we offer an open-source library that analyzes low-lying STZs in computer glasses and in laboratory materials such as dense colloidal suspensions for which the harmonic approximation is accessible. Our results constitute a physically motivated methodological advancement towards characterizing mechanical disorder in glasses, and understanding how they yield. 
\end{abstract}

\maketitle

\emph{Introduction.}---It has been known since the seminal works of Spaepen and Argon in the late 1970s \cite{spaepen_1977,argon_st} that plastic flow in amorphous solids proceeds via stress-induced localized rearrangements of small clusters of particles. Those rearrangements, their collective dissipative dynamics and spatiotemporal correlations give rise to many emergent phenomena such as plastic strain localization \cite{falk_prl_2005,falk_prb_2006,lisa_strain_localization_pre_2007,talamali_2012}, shear banding \cite{Ozawa6656,fsp,falk_pre_2018}, system spanning avalanches of plastic activity \cite{lemaitre2004_avalanches,lemaitre2006_avalanches,steady_states_with_jacques,talamali_pre_2011,salerno_robbins,yielding_pnas_mw_2014, Ezequiel_prl_2016, sri_nat_com_2017}, and macroscopic yielding \cite{Schuh_review_2007,Falk2011,falk_review_2016}. 

A first-principles understanding of these emergent phenomena calls for the identification and statistical quantification of the microstructural entities that constitute the precursors of stress-induced dissipative rearrangements in amorphous solids. Those precursors and their micromechanical nature were envisioned by Falk and Langer two decades ago \cite{falk_langer_stz}, and subsequently coined \emph{Shear Transformation Zones} (STZs). Phenomenological theories \cite{falk_langer_stz,lisa_strain_localization_pre_2007,eran_jim_prl_2011,eran_jim_soft_matter_2013,eran_jmp_2014} and several variants of elastoplastic lattice models \cite{Roux_elasto_plastic_model_prl_2002, yielding_pnas_mw_2014,Ozawa6656,barrat_elasto_plastic_model_review_2018} were since put forward, building on the premise that a population of STZs is encoded in a glass's structure, and serves as the key vehicle for plastic deformation and macroscopic yielding. 

Substantial computational research efforts have been dedicated to the search for structural indicators that serve as faithful representatives of STZs, see Ref.~\cite{david_huge_collaboration} for an extensive review of those efforts. In parallel, micromechanical theories of elastoplastic instabilities, formulated within the potential energy landscape picture \cite{Goldstein1969}, have been put forward, both in the harmonic \cite{Malandro_Lacks,lemaitre2004,lemaitre2006_avalanches} and anharmonic \cite{barriers_lacks_maloney, micromechanics2016,episode_1_2020} regimes. In these formulations, STZs are represented by destabilizing modes (putative displacement fields about the mechanical equilibrium state) whose associated energies vanish continuously upon approaching the onset of elastoplastic instabilities under external deformation~\cite{micromechanics2016}. Using harmonic modes to detect STZs is a natural starting point, as they are simple and efficient to calculate.

One clear limitation of the harmonic formulation of elastoplastic instabilities is the tendency of soft, quasilocalized vibrational modes --- that destabilize under external deformations --- to hybridize with other low-frequency modes, primarily phononic \cite{micromechanics2016, SciPost2016, phonon_widths}, but also quasilocalized \cite{episode_1_2020}. Consequently, the utility of harmonic analyses in exposing \emph{quantitative} information regarding plastic instabilities is system-size dependent; in particular, only at strains of order $\lesssim\! L^{-4}$ away from plastic instabilities (in systems of linear size $L$), does quantitative micromechanical information regarding the imminent instability become available by studying the lowest vibrational mode of a glass~\cite{micromechanics2016}. At larger strains away from instabilities, hybridizations spoil said information, as demonstrated in Fig.~\ref{fig:fig1} below.

\begin{figure*}[!bt]
\centering
\includegraphics[width = 1.0\textwidth]{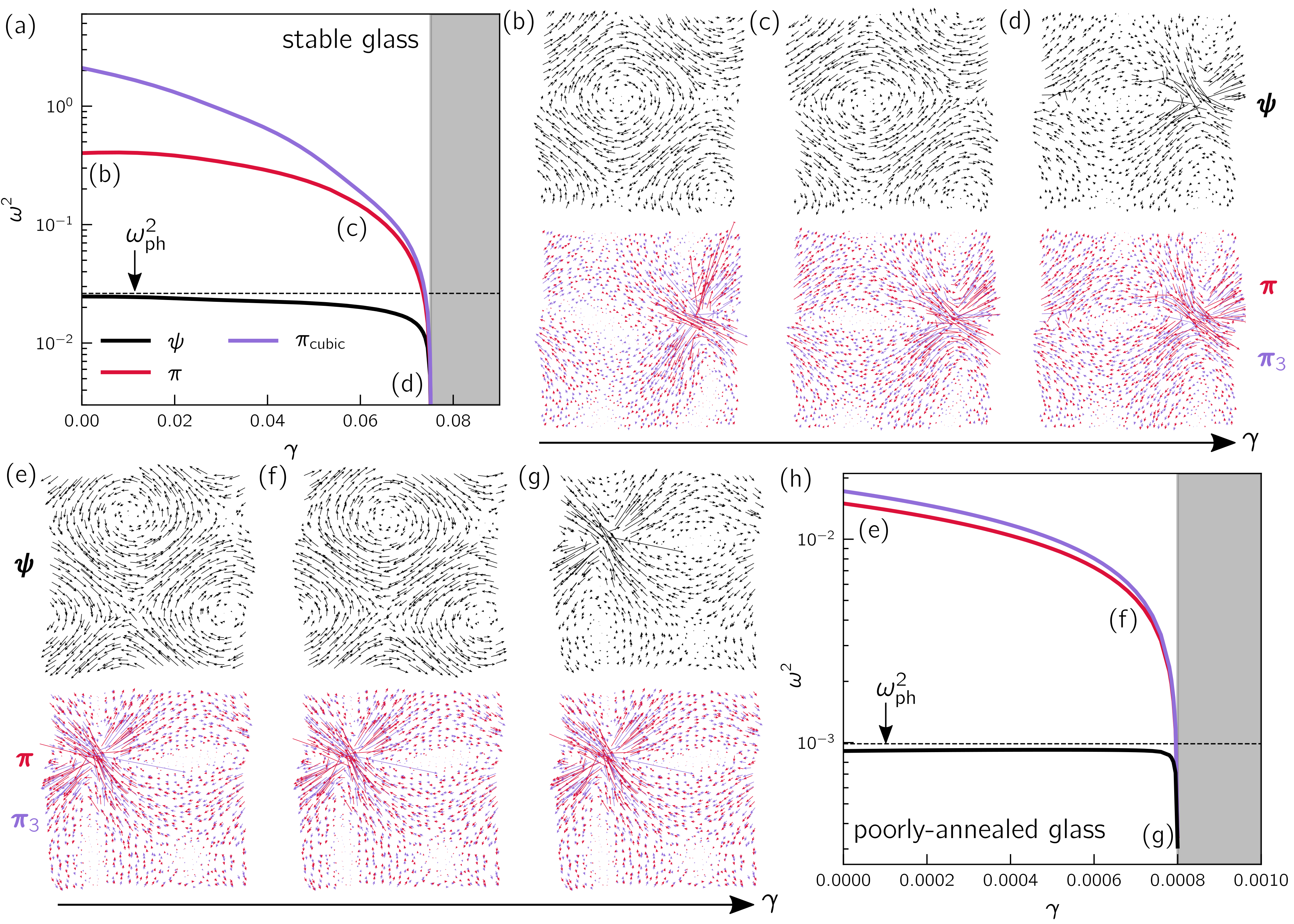}
\caption{\footnotesize Pseudo-harmonic modes represent STZs in glasses, ranging from ultrastable ($T_p\!=\!0.2$, panels (a)-(d)) to poorly-annealed ($T_p\!=\!0.7$, panels (e)-(h)). (a) The squared frequencies of a destabilizing PHM $\piv$ (solid red line), cubic mode  $\piv_3$ (solid purple line), and vibrational mode $\psiv$ (solid black line) of an ultrastable glass of $N\!=\!10$K particles in two dimensions, subjected to athermal quasistatic shear (see details in the Appendix~\ref{ap:sec1}). The horizontal dashed line indicates the first shear wave frequency $\omega_{\rm ph}\!=\!2\pi c_s/L$, with $L$ and $c_s(T_p)$ denoting the box length and (glass-history-dependent) shear-wave speed, respectively. Panels (b)-(d) show $\psiv$ (top row), $\piv$ and  $\piv_3$ (2nd row) at the strains indicated in panel (a), corresponding to strain \emph{differences} away from the instability of $\Delta\gamma\!=\!10^{-1}$, $10^{-2}$, and $10^{-5}$, from left to right. These data show that the firstly-activated STZ under shear is present in the as-cast glass, in the form of a PHM. Panels (e)-(h) are the same as (a)-(d), but measured in a poorly annealed glass of $N\!=\!40$K particles, with $\Delta\gamma\!=\!8\!\times\!10^{-4}, 2\!\times\!10^{-4}$, and $10^{-5}$, from left to right.}
\label{fig:fig1}
\end{figure*}

A potential solution to some of the obstacles posed by hybridization issues seen in harmonic frameworks was recently put forward, in the form of a nonlinear micromechanical framework \cite{plastic_modes_prerc,micromechanics2016,SciPost2016,episode_1_2020}. Within this framework, the microstructural entities that constitute the precursors of elastoplastic instabilities are (normalized) displacement fields $\piv_3$, coined \emph{plastic} modes or \emph{cubic} modes, which are defined as solutions to the algebraic equation
\begin{equation}\label{cubic_modes_definition}
\frac{\partial ^2U}{\partial\xv\partial\xv}\cdot{\piv_3} = \frac{\frac{\partial ^2U}{\partial\xv\partial\xv}:\piv_3\piv_3}{\frac{\partial ^3U}{\partial\xv\partial\xv\partial\xv}\tripleCdot\piv_3\piv_3\piv_3} \frac{\partial ^3U}{\partial\xv\partial\xv\partial\xv}:\piv_3\piv_3\,,
\end{equation}
where $U(\xv)$ is the potential energy that depends on coordinates $\xv$, and $:,:\!\cdot$ represent double and triple contractions, respectively. Cubic modes $\piv_3$ were shown to feature nontrivial statistical \cite{episode_1_2020} and micromechanical \cite{micromechanics2016} properties, and can be considered as one of the most informative representatives of STZs, as discussed in more detail in the Appendix~\ref{cubevspseudo}.

In parallel to the apparent utility (see, e.g., Refs.~\cite{cge_paper,pinching_pnas,episode_1_2020}) of the nonlinear micromechanical framework within which cubic modes are defined, its general applicability to computer glass models is limited: in several well-studied models, higher-order ($\ge$ 3rd) spatial derivatives of the potential energy --- necessary for the computation of cubic modes, as evident by Eq.~(\ref{cubic_modes_definition}) --- are either impossible to evaluate, e.g.~in hard sphere glasses, cumbersome to evaluate, e.g.~in the Stillinger-Weber model \cite{Stillinger_Weber} that features a 3-body interaction term, or singular by construction, e.g.~in Hertzian spheres near the unjamming point~\cite{ohern2003,liu_review,van_hecke_review}. 

Here we introduce a simple, alternative definition of soft, quasilocalized modes --- referred to in what follows as \emph{pseudo-harmonic modes} (PHMs) --- and directly demonstrate that they faithfully represent STZs. A key feature of PHMs' is that their definition relies solely on the availability of the harmonic approximation of the potential (or free) energy --- in the form of the Hessian matrix $\calBold{H}\!\equiv\!\partial^2U/\partial\xv\partial\xv$ --- and not on higher order derivatives, as some previous definitions of STZs do \cite{plastic_modes_prerc, SciPost2016, lte_pnas, zohar_prerc, episode_1_2020}. As demonstrated below, the PHM framework is broadly applicable, straightforward, and computationally efficient. We further provide physical arguments that motivate our definition of PHMs, and show that in the zero frequency limit, the frequencies associated with PHMs converge to those associated with the \emph{softest} nonphononic vibrational modes. Finally, we offer a software library~\cite{package} that calculates low-energy STZs via the presented framework, for any given Hessian of a glass in mechanical equilibrium. 

\newpage

\emph{Pseudo-harmonic modes}.---PHMs are putative displacement fields $\piv$ about a mechanical equilibrium state, for which the cost function \cite{episode_1_2020}
\begin{equation}\label{cost_function}
{\cal C}(\zv) = \frac{\big(\calBold{H}:\zv\zv\big)^2}{\sum\limits_{\mbox{\tiny $\langle i,\! j\rangle$}}\big(\zv_{ij}\cdot\zv_{ij}\big)^2}\,,
\end{equation}
assumes local minima, i.e.~they solve
\begin{equation}\label{phm_definition}
\frac{\partial {\cal C}}{\partial \zv}\bigg|_{\zv=\mbox{\footnotesize$\piv$}}=\zerovector\,.
\end{equation}
Here $\zv_{ij}\!\equiv\!\zv_j\!-\!\zv_i$, and the sum in Eq.~(\ref{cost_function}) runs over all pairs $\langle i,\! j\rangle$ of interacting particles \cite{footnote}. It is apparent by examining Eq.~(\ref{cost_function}) that PHMs are accessible in any system whose Hessian matrix $\calBold{H}$ is available, which is a major strength of our approach, demonstrated further below.

Why do PHMs $\piv$ --- for which the cost function ${\cal C}(\zv)$ given by Eq.~(\ref{cost_function}) assumes local minima --- constitute faithful descriptors of STZs? This point is demonstrated explicitly in Fig.~\ref{fig:fig1}, but can be argued for as follows; when evaluated at local minima $\piv$ of the cost function ${\cal C}(\zv)$, ${\cal C}(\piv)$'s numerator is expected to be small, and its denominator -- large. The numerator of ${\cal C}(\zv)$ describes the square of (twice) the energy associated with $\zv$ (assuming harmonicity), therefore $\piv$ will generally represent a low-frequency mode. ${\cal C}(\zv)$'s denominator $\sum\nolimits_{\mbox{\tiny $\langle i,\! j\rangle$}}(\zv_{ij}\cdot\zv_{ij})^2$ can be argued to $(i)$ scale as $k^4$ for waves of wave number $k$ -- and therefore strongly suppress long wavelength phononic modes, and $(ii)$ be inversely proportional to $\zv$'s participation ratio $e(\zv)\!\equiv\!(N\sum\nolimits_i \mbox{\small (}\zv_i\cdot\zv_i\mbox{\small )}^2)^{-1}$ (demonstrated in Appendix~\ref{denominator}) -- and is therefore larger (smaller) for more (less) localized modes. These features of ${\cal C}(\zv)$ explain why PHMs $\piv$ that represent its local minima are generally both soft and quasilocalized modes, as required in order to constitute STZs.

Solutions $\piv$ to Eq.~(\ref{phm_definition}) can be readily obtained in two ways described next.
\begin{enumerate}
\item Starting with an initial guess $\piv^{(0)}$, repeatedly apply the mapping 
\begin{equation}
\calBold{F}(\piv) = \frac{\calBold{H}^{-1}\cdot\zetav(\piv)}{\sqrt{\zetav(\piv)\cdot\calBold{H}^{-2}\cdot\zetav(\piv)}}\,,
\end{equation}
where 
\begin{equation}
\zetav_k(\piv)\!\equiv\!\sum\nolimits_{\mbox{\tiny $\langle i,\! j\rangle$}}\!\!(\delta_{jk}\!-\delta_{ik})(\piv_{ij}\cdot\piv_{ij})\piv_{ij}\,,
\end{equation}
until $\calBold{F}(\piv)\!\simeq\!\piv$, which can be shown to be equivalent to Eq.~(\ref{phm_definition}). 
\item Starting with an initial guess $\piv^{(0)}$, minimize the cost function ${\cal C}(\zv)$ given by Eq.~(\ref{cost_function}) to obtain a PHM $\piv$.
\end{enumerate}
The iterative scheme (1) has the advantage of being parameter free, and only requires solving a set of linear equations (at each iteration). The minimization scheme (2) is computationally more efficient, however it inherits the disadvantage of nonlinear minimization algorithms, which require the choice of problem-dependent parameters.

An example of a PHM calculated in a two-dimensional computer glass subjected to athermal, quasistatic (AQS) shear deformation is shown in Fig.~\ref{fig:fig1}. We show that the harmonic and nonlinear descriptions of the elastoplastic instability agree as the shear strain $\gamma$ approaches the instability strain $\gamma_c$. At strains $\gtrsim\! L^{-4}$ away from $\gamma_c$, the harmonic description breaks down due to phonon hybridizations~\cite{micromechanics2016}, while the nonlinear description persists to reflect the geometry and locus of the imminent instability, up to large $\Delta\gamma=\gamma_c\!-\!\gamma\simeq7\%$ (in the example of Fig.~\ref{fig:fig1}). Moreover, PHMs closely resemble cubic modes along the whole elastic branch. Cubic modes have a higher stiffness because the third-order coefficient of the expansion of the potential energy is very sensitive to the structure of the mode, as discussed in detail in Ref.~\cite{episode_1_2020}.

\begin{figure}[!ht]
\centering
\includegraphics[width = 0.5\textwidth]{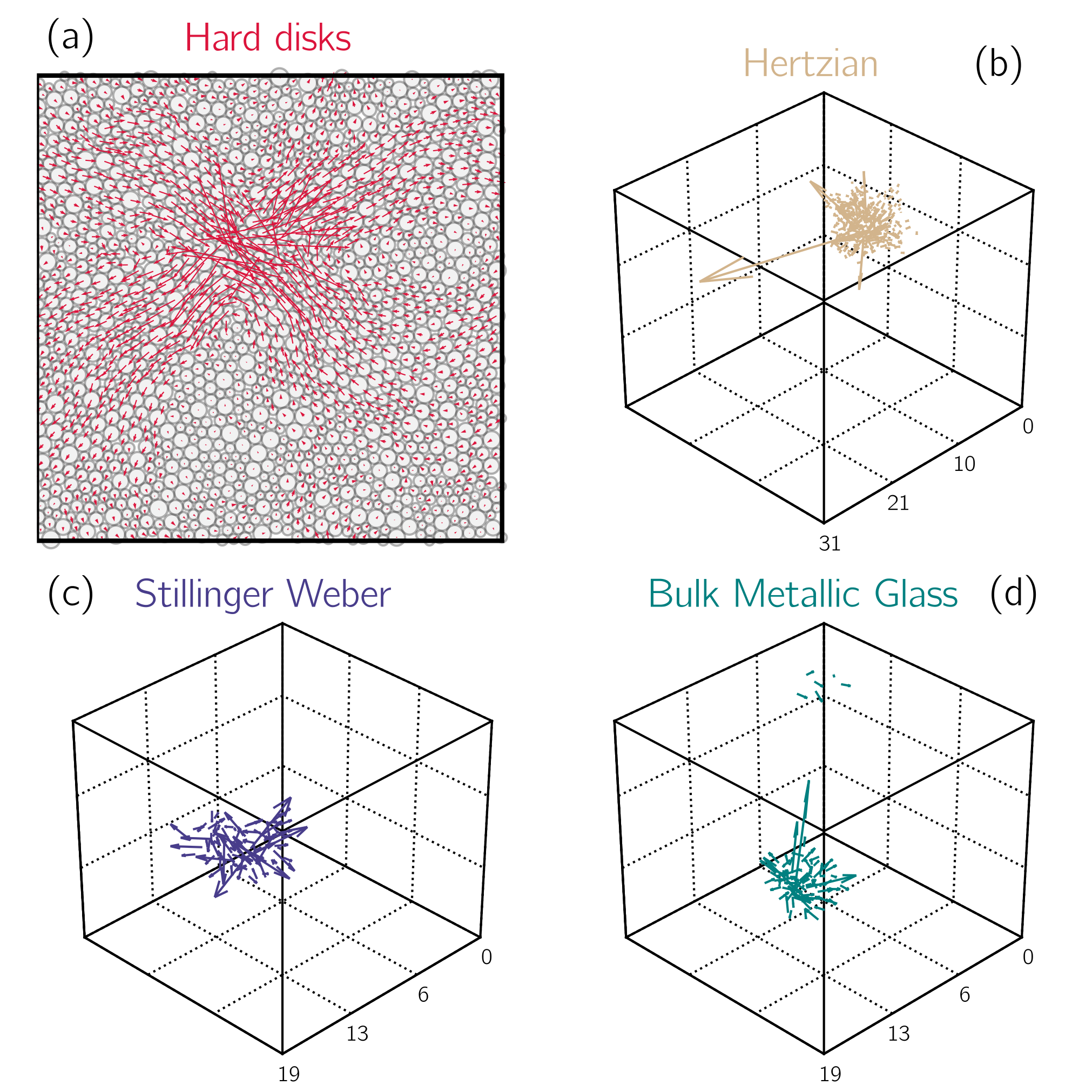}
\caption{\footnotesize Pseudo Harmonic Modes (PHMs) in various computer glasses: (a) a hard-disk glass, (b) a glass of Hertzian soft spheres, (c) a Stillinger-Weber tetrahedral glass with 3-body interactions~\cite{Stillinger_Weber}, and (d) a CuZr Bulk Metallic Glass~\cite{cheng2009atomic}.}
\label{fig:fig2}
\end{figure}

\emph{General applicability of PHMs}. In Fig.~\ref{fig:fig2} we present PHMs calculated for various computer glass models~\ref{ap:sec1} for which extracting soft, quasilocalized modes using the anharmonic micromechanical framework presented in \cite{plastic_modes_prerc, SciPost2016, episode_1_2020} is either very difficult or impossible. In particular, we show a PHM found in as-cast (not deformed) glasses of (a) hard spheres, (b) Hertzian soft spheres, (c) the Stillinger-Weber model \cite{Stillinger_Weber}, and finally (d) a Copper-Zirconium bulk metallic glass (BMG) model \cite{cheng2009atomic}. Details about the models and calculations are presented in the Appendix\ref{ap:sec1}. 

\emph{Convergence to nonphononic harmonic modes}. A stringent benchmark for various definitions of soft, quasilocalized modes is the degree of agreement between their structural and energetic properties, to those associated with nonphononic harmonic (vibrational) modes representing the same soft spots in the material. Detailed discussions regarding this benchmarking, and its implications, can be found in~\cite{episode_1_2020}. 

\begin{figure}[!ht]
\centering
\includegraphics[width = 0.5\textwidth]{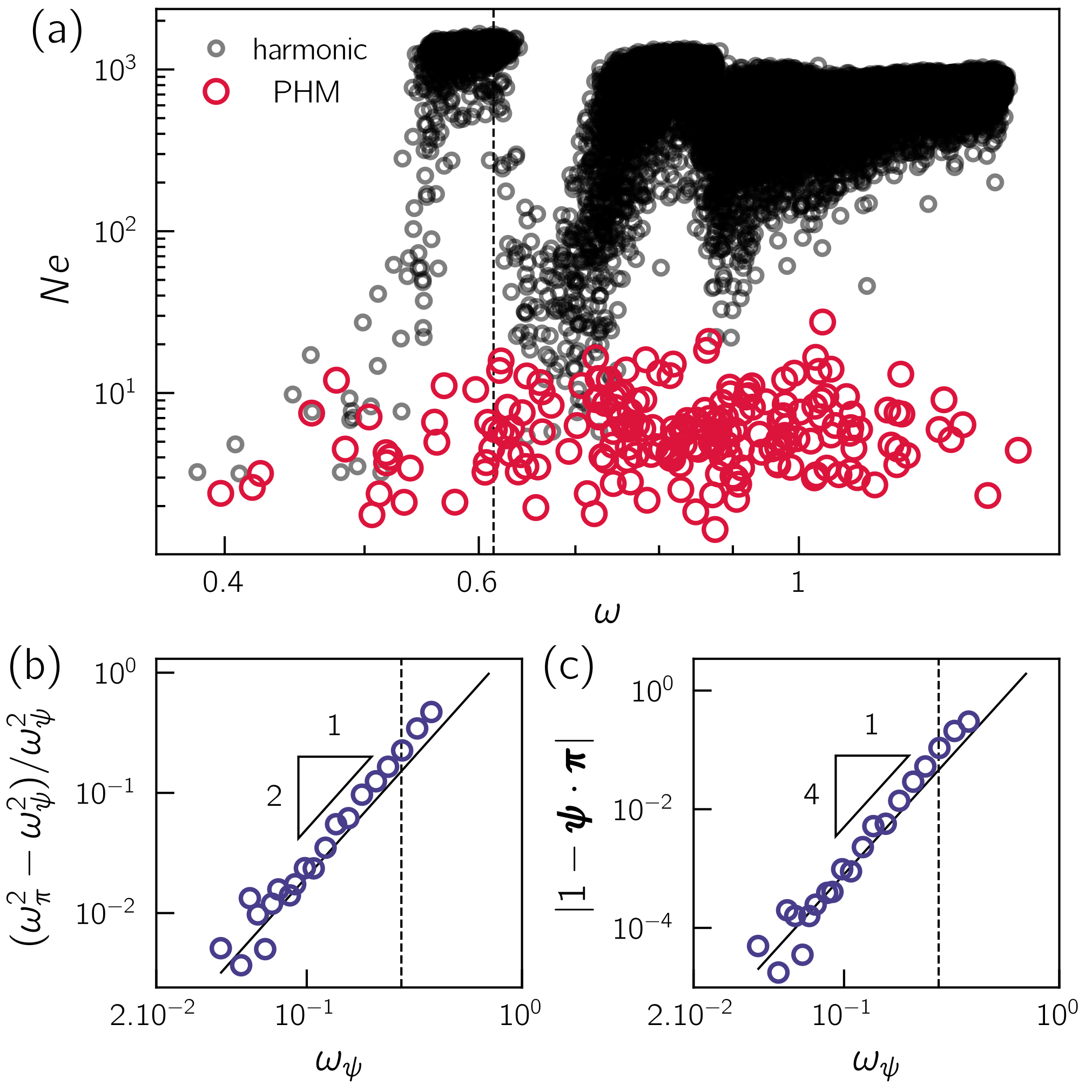}
\caption{\footnotesize Pseudo Harmonic Modes (PHMs) versus harmonic vibrational modes: (a) Scaled participation ratio $Ne$ as a function of frequency $\omega$, calculated in an ensemble of computer glasses in three dimensions (see Appendix~\ref{ap:sec1}). The convergence of the energy (panel (b)) and structure (panel (c)) of PHMs to those of their ancestral harmonic modes. In every panel the dashed line indicates the lowest phonon frequency.}
\label{fig:fig3}
\end{figure}

Here we build an ensemble of PHMs, one for each glassy sample (see Appendix\ref{ap:sec1}); we do this by setting $\piv^{(0)}$ to be the softest \emph{harmonic} mode $\psiv$ in a given glass, which has an energy $\omega_{\psi}^2$ (setting units such that all masses are unity). We then map $\piv^{(0)}\!=\!\psiv$ to a PHM $\piv$ with energy $\omega_{\pi}^2$ using either of the two methods described above (the result is independent of this choice). In Fig.~\ref{fig:fig3}a we compare the obtained solutions $\piv$ with low-frequency harmonic modes by scatter-plotting each mode's localization --- as captured by the scaled participation ratio $Ne$ --- versus its associated frequency. We see that PHMs remain localized irrespectively of their frequency, whereas harmonic modes show a strong hybridization with plane waves above the lowest phonon-frequency $\omega_{\rm ph}\!=\!2\pi c_s/L$ \cite{ikeda_pnas,phonon_widths}, where $c_s$ denotes the shear wave speed. Finally, we show in Fig.~\ref{fig:fig3}(b)-(c) the average relative energy differences $(\omega_{\pi}^2\! - \omega_{\psi}^2)/\omega_{\psi}^2$ and the average overlaps $1\!-\!|\piv\cdot\psiv|$, as a function of the harmonic modes' frequencies $\omega_{\psi}$, and binned over those same frequencies. We find that as $\omega_{\psi}\!\!\to\!0$ solutions $\piv$ converge both energetically and structurally to harmonic modes $\psiv$. The implications of this convergence are discussed~below.

To further demonstrate the veracity of PHMs as true plastic defects mediating STZs, we compare the map of the residual strength --- local yield stress --- measured with the frozen matrix method~\cite{patinet2016connecting,barbot2018local} with the location of PHMs at various plastic events. Here, we map at each shear transformation the triggering critical mode $\piv^{(0)}\!=\!\psiv_c$ to a PHM $\piv$ computed from the as-cast ($\gamma=0$) cost function. Surprisingly, we find that all plastic events can be traceback to PHMs at zero strain. This result firmly establishes that regions with low residual strength emanate from the presence of soft quasi-localized excitations.

\begin{figure}[!ht]
\centering
\includegraphics[width = 0.5\textwidth]{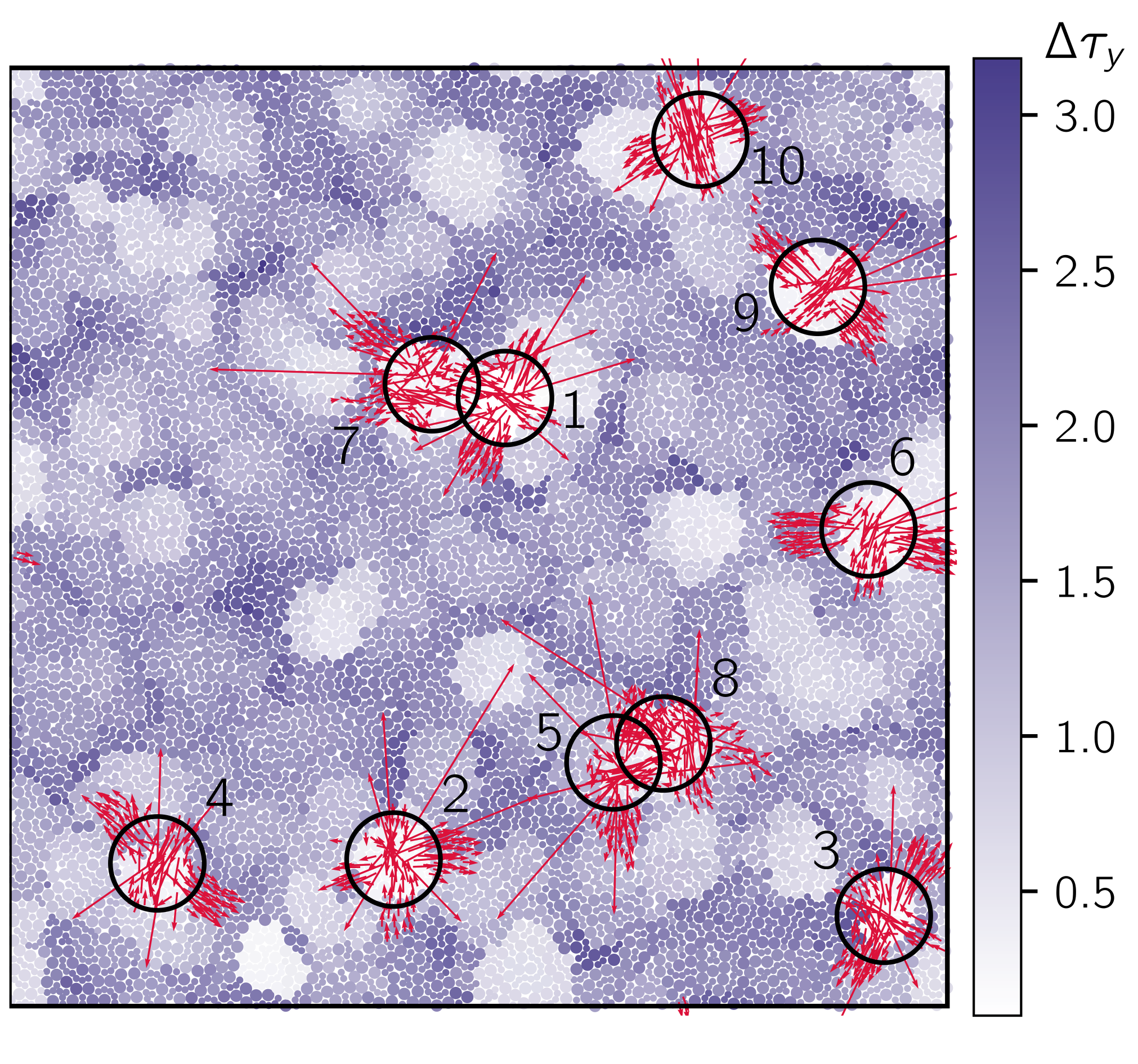}
\caption{\footnotesize Residual plastic strength map of a well-annealed Lennard-Jones binary glass at zero strain $\gamma=0$ (details about the model can be found in Ref.~\cite{barbot2018local}). Black circles indicate the loci of the 10 first plastic instabilities. PHMs are extracted from the as-cast configurations by mapping the critical mode at each plastic event.}
\label{fig:fig4}
\end{figure}

\emph{Summary and outlook}. Revealing the micromechanical entities that carry plastic flow in amorphous solids --- the shear transformation zones (STZs) --- are key to understanding these materials' failure mechanisms. To this aim, various methods designed to identify a population of STZs in glassy solids via a \emph{harmonic} analysis of their potential energy have been put forward~\cite{widmer2008irreversible,xu2010anharmonic,tanguy2010,manning2011,rottler_normal_modes,manning_defects}. These methods feature appreciable degrees of success in predicting plastic flow \cite{david_huge_collaboration}, including in experimental setups \cite{chen2011measurement}. However, they typically do not provide a micromechanical characterization of a single, isolated STZ, in terms of its energy, orientation, and coupling to external loads, and are further hindered by hybridizations that typically occur between different vibrational modes.

In this Letter we have introduced a simple and generally-applicable micromechanical definition of STZs. These objects are referred to here as \emph{pseudo-harmonic modes} (PHMs) because they depend solely on the harmonic approximation of the potential (or free) energy (in the form of its Hessian matrix). We show that PHMs can be calculated in a variety of model systems in which other soft, nonlinear excitations are either inaccessible or cumbersome to obtain. We demonstrate that PHMs are good indicators of elastoplastic instabilities in the athermal, quasistatic limit, and show that their associated energies in as-cast glassy samples converge to nonphononic, harmonic modes' energies, in the low-energy limit. Finally, an open-source library is offered~\cite{package}, that calculates PHMs given a glass's Hessian matrix.

The convergence of the spatial structure and energies of PHMs and of low-frequency, nonphononic harmonic modes, suggests that the formers' frequency distribution follows the same asymptotic $\sim\!\omega^4$ law, which is universally featured by the nonphononic density of states of structural glasses quenched from a melt \cite{modes_prl_2016,modes_prl_2018,modes_prl_2020}.

The ability to extract the precise displacement field associated with a single STZ, including very far (in strain) from its eventual instability, and exclusively using the harmonic approximation of the energy, opens up a wide range of new analyses in computer glasses and some experimental systems. For example, our method could be used to systematically quantify the properties of STZs in a wide variety of glass models (available e.g.~in LAMMPS \cite{plimpton1993fast}), as a function of composition or material preparation \cite{pan2008experimental}. In addition, our method could be used to quantify the orientation and density of STZs, and study their evolution as a function of applied shear strain \cite{xu2019atomic}, which would place strong constraints on continuum models for plasticity in amorphous solids \cite{falk_langer_stz,lisa_strain_localization_pre_2007,eran_jim_prl_2011,eran_jim_soft_matter_2013,eran_jmp_2014}. 

We note finally that while some interesting insights into glass physics have been obtained \cite{cge_paper,pinching_pnas,episode_1_2020} by investigating soft anharmonic excitations using existing frameworks \cite{SciPost2016,episode_1_2020}, an algorithm to detect all such excitations in a given glassy sample is still under development \cite{in_preparation}. The ideas presented here might also find utility in saddle-point search algorithms such as the activation-relaxation technique~\cite{art_2001}, or in searches for two-level system in computer glasses \cite{Ruocco_1999}.

A readily usable PYTHON package to detect soft spots in structural glasses is available in Ref.~\cite{package}. It includes documentation and a minimal example.

\emph{Acknowledgements.} We warmly thank Talya Vaknin and Eran Bouchbinder for discussions and numerical support. We are very grateful to Sylvain Patinet for providing us the residual plastic strength data. We further thank Corrado Rainone and Karina Gonz\'alez-L\'opez for their comments on the manuscript. We are grateful for the support of the Simons Foundation for the `Cracking the Glass Problem' Collaboration Awards No.~348126 to Sid Nagel (D.~R.), No.~454947 (M.~L.~M.), and Simons investigator Grant No.446222 (J.~A.~G). E.~L.~acknowledges support from the Netherlands Organisation for Scientific Research (NWO) (Vidi Grant No.~680-47-554/3259). This work was carried out on the Dutch national e-infrastructure with the support of SURF Cooperative.

\appendix

\section{Computer glass models}
\label{ap:sec1}
\subsection{Inverse Power Law}
The results shown in Figs.~1 and~3 in the main text are for polydisperse soft spheres in two and three dimensions, respectively, interacting via an inverse power law potential. A detailed description of this model is provided in Ref.~\cite{scattering_jcp}. We utilize SWAP Monte Carlo~\cite{berthier_swap_hard_spheres, LB_swap_prx} (MC) to prepare glasses with various degrees of stability. The later is controlled by the parent temperature $T_p$ of the equilibrium states from which our glasses were instantaneously quenched. Finally, we quench our configurations to zero temperature via an energy minimization using a conjugate gradient algorithm \cite{macopt_cg}.

Athermal quasistatic shear deformation~\cite{Malandro_Lacks} is performed using Lees-Edwards periodic boundary conditions \cite{allen1989computer} with a strain step $\delta\gamma\!=\!10^{-5}$, which is at least one order of magnitude lower than the typical strain between subsequent plastic instabilities.

\subsection{Hard disks}
\begin{figure}[!ht]
\centering
\includegraphics[width = 0.5\textwidth]{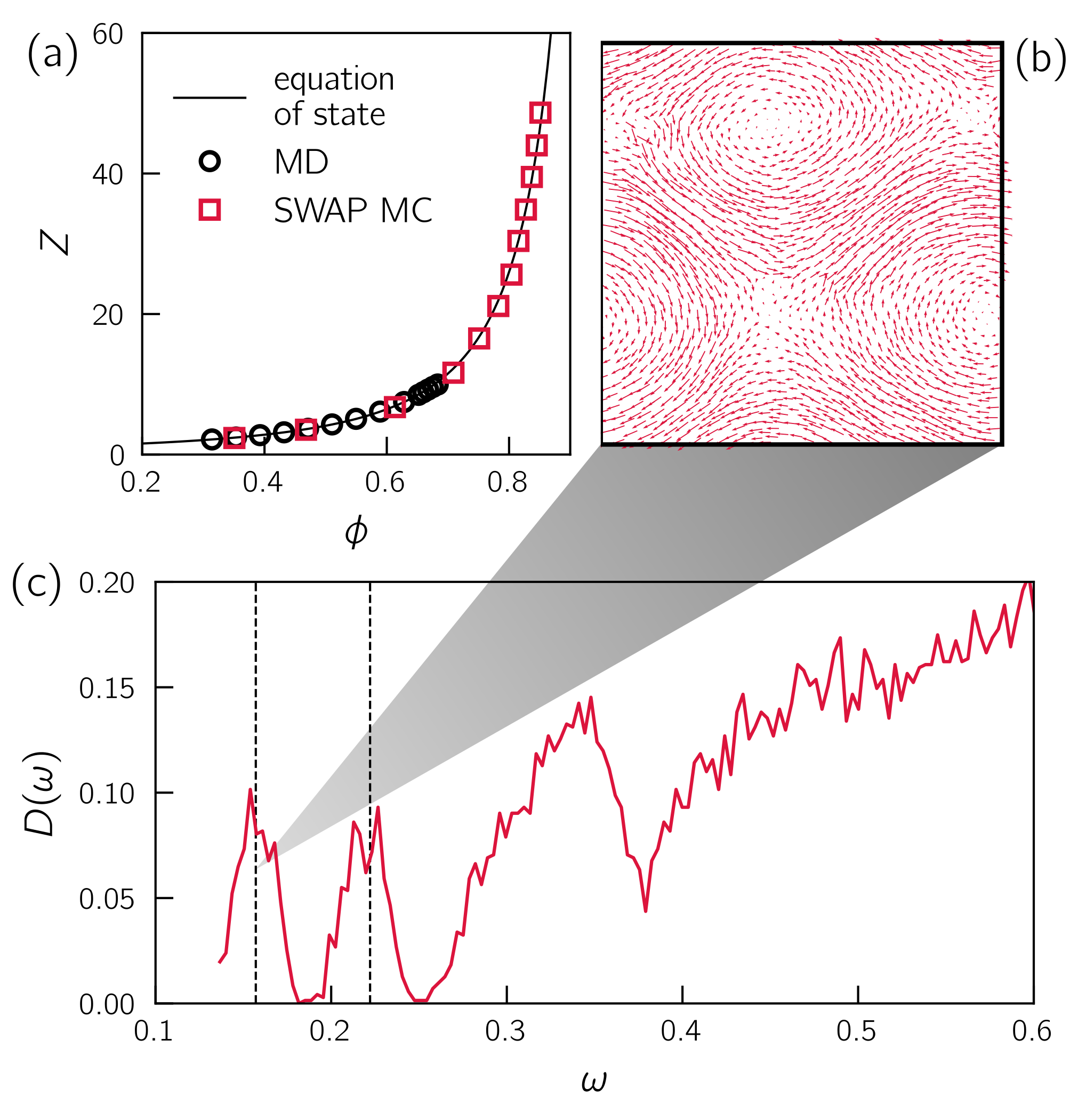}
\caption{\footnotesize (a) Compressibility $Z=P/(\rho k_BT)$ as a function of the packing fraction $\phi$. The solid line is the equation of state taken from Ref.~\cite{mulero2009equation}, empty black symbols are Molecular Dynamics results from Ref.~\cite{kolafa2006simulation}~, and red empty squares are SWAP Monte Carlo simulations performed in the NPT ensemble. (b) Typical low-frequency transverse phonon. (c) Density of states in hard-disk glasses with $N=1600$ prepared at $Z=35$. The vertical dashed lines indicate the frequencies of the first and second shear waves.}
\label{fig:apfig1}
\end{figure}

We prepare dense equilibrium polydisperse hard-disk configurations with $N=1600$ particles using SWAP MC. We choose the same continuous polydispersity as proposed in Ref.~\cite{berthier_swap_hard_spheres} with diameter distribution $P(\sigma)\!\sim\!\sigma^{-3}$ from $\sigma_m\!=\!1$ to $\sigma_M\!=\!2.2$. This choice results in a high polydispersity $\Delta\!=\!\sqrt{\overline{\sigma^2}-\overline{\sigma}^2}/\overline{\sigma}\!\simeq\!22\%$ and as a result one avoids crystallization. Throughout our simulations lengths are in units of $\overline{\sigma}$. Initial configurations are prepared via minimization of harmonic soft spheres at a high packing fraction $\phi\simeq0.82-0.84$. We then perform a MC run in the NPT ensemble where the box edge is allowed to fluctuate by around $3\%$ of its length. In Fig.~\ref{fig:apfig1}a, we show the compressibility $Z=P/(\rho k_BT)$ as a function of the packing fraction $\phi$. As previously demonstrated~\cite{berthier_swap_hard_spheres}, SWAP MC enables the preparation of equilibrium configurations far above the glass transition packing fraction $\phi_g\!\simeq\!0.7$.

Having generated equilibrium dense hard-disk configurations, we perform MC simulations in the NVT ensemble to compute the positions covariance matrix
\begin{equation}
\calBold{C}_{ij}=\langle \uv_i(t) \uv_j(t)\rangle_T,
\end{equation}
with the particle displacement vector $\uv_i=\xv_i-\langle\xv_i\rangle$ from its average position. Note that during this procedure we have removed possible drift due to the motion of the center of mass of the simulation box. The Hessian matrix of the system follows from the equality~\cite{henkes2012extracting}
\begin{equation}
\calBold{H}_{ij}=k_{\mbox{\scriptsize B}}T\calBold{C}^{-1}_{ij}\,,
\end{equation}
with $k_{\mbox{\scriptsize B}}$ the Boltzman constant and where particle masses are equal and set to unity. Full diagonalizations are performed using the LAPACK library. In Fig.~\ref{fig:apfig1}b, we show a typical phonon found in deeply equilibrated hard-disk glasses. Furthermore, we have generated an ensemble of $128$ configurations for $Z\!\simeq\!35$ and computed the density of states $D(\omega)$ with frequency $\omega$ (see Fig.~\ref{fig:apfig1}c). As discussed in depth in Ref.~\cite{phonon_widths}, we observe a finite size regime at low-frequencies with distinct phonon bands followed by a continuous ``phonon sea" at higher frequencies. Consistent with the degeneracy level of phonon bands, we have found $4$ modes in both the first and second bands.

The extraction of PHMs requires the contact network of pairs $\{ij\}$ to be able to compute the denominator of the cost function defined in Eq.~(2) of the main text. As contacts are not accessible in hard particles, we have chosen the $6$ closest neighbors of each particles (first peak of the radial distribution function). We have checked that the energy and structure of the modes are not affected in any appreciable manner by this choice.

\subsection{Other models}

A complete description of the Hertzian, Stillinger-Weber, and CuZr BMG models can be found in Ref.~\cite{modes_prl_2020}. The Lennard-Jones configuration used in the main text for the residual palstic strength map is the same one as shown in Ref.~\cite{patinet2016connecting}.

\begin{figure}[!h]
\centering
\includegraphics[width = 0.5\textwidth]{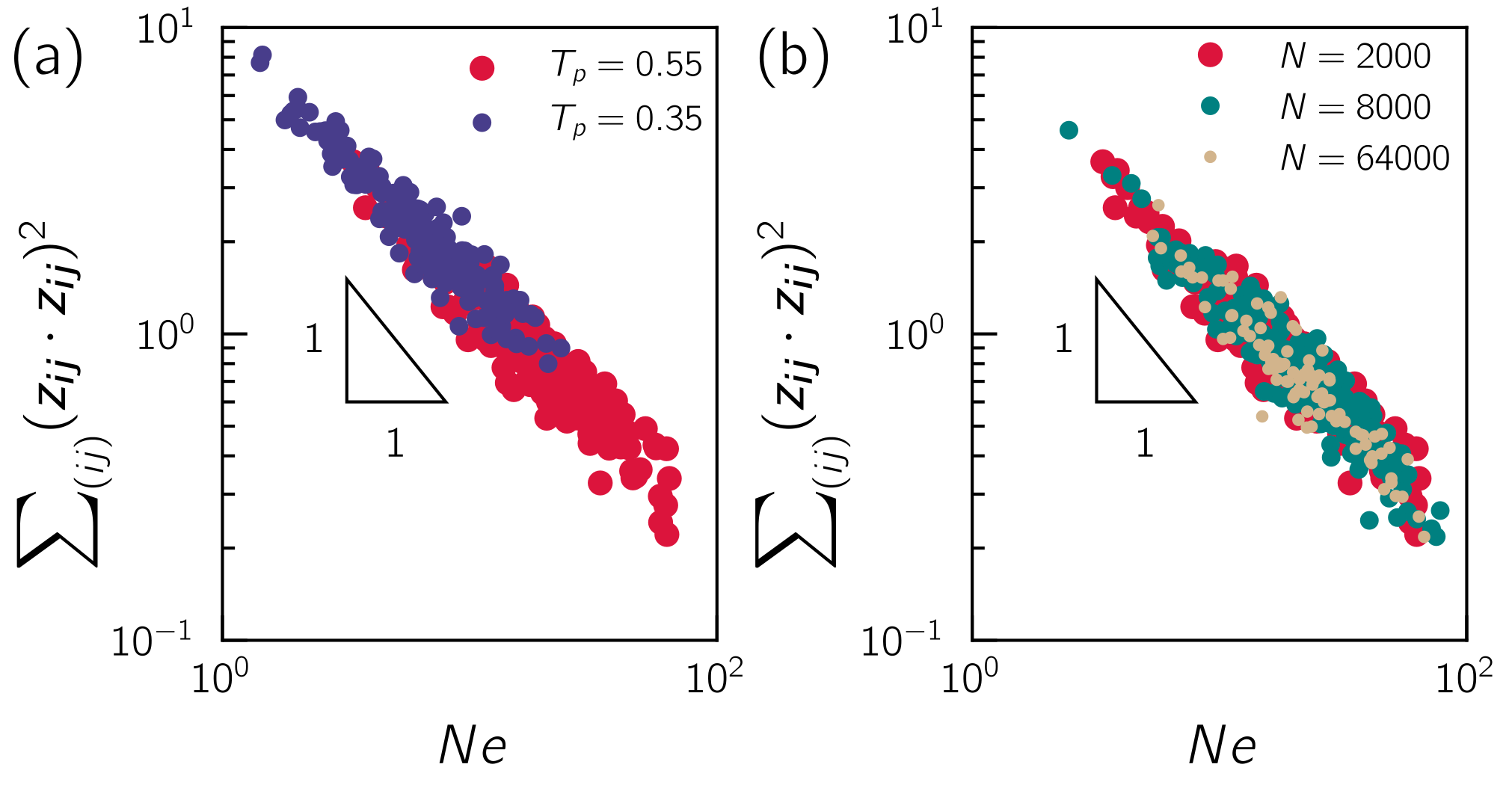}
\caption{\footnotesize (a) The demoninator of the cost function ${\cal C}(\zv)$, evaluated for PHMs $\mathBold{\pi}$, and plotted against the scaled participation ratio $Ne$, measured for glassy samples of $N\!=\!2000$ particles of a polydispersed soft spheres model~\cite{pinching_pnas}. Here the degree of localization is varied by considering glasses quenched from different parent temperatures $T_p$ as indicated by the legend. (b) Same as panel (a), this time for a single parent temperature $T_p\!=\!0.55$ (in simulation units), but different system sizes.}
\label{fig:apfig2}
\end{figure}

\vspace{-0.5cm}

\section{Cost function denominator}
\label{denominator}
In Fig.~\ref{fig:apfig2} we present data that demonstrates that the denominator of the cost function ${\cal C}(\zv)$ (see Eq.~2) in the main text) follows $\sum\nolimits_{\mbox{\tiny $\langle i,\! j\rangle$}}(\zv_{ij}\cdot\zv_{ij})^2 \sim 1/\big(Ne(\zv))$, where $e(\zv)\!\equiv\!(N\sum\nolimits_i \mbox{\small (}\zv_i\cdot\zv_i\mbox{\small )}^2)^{-1}$ is the conventional participation ratio, $N$ is the system size, and $\zv$ denotes a normalized ($\zv\cdot\zv\!=\!1$) putative displacement field. This means that the denominator of the cost function ${\cal C}(\zv)$ is larger (smaller) for more (less) localized modes, therefore promoting localization of its minima $\mathBold{\pi}$ (the PHMs). In Fig.~\ref{fig:apfig2} we vary both the system size, and the degree of modes' localization -- the latter is known to be affected by the degree of supercooling, captured by their parent temperature $T_p$~\cite{pinching_pnas}.

\begin{figure}[!h]
\centering
\includegraphics[width = 0.5\textwidth]{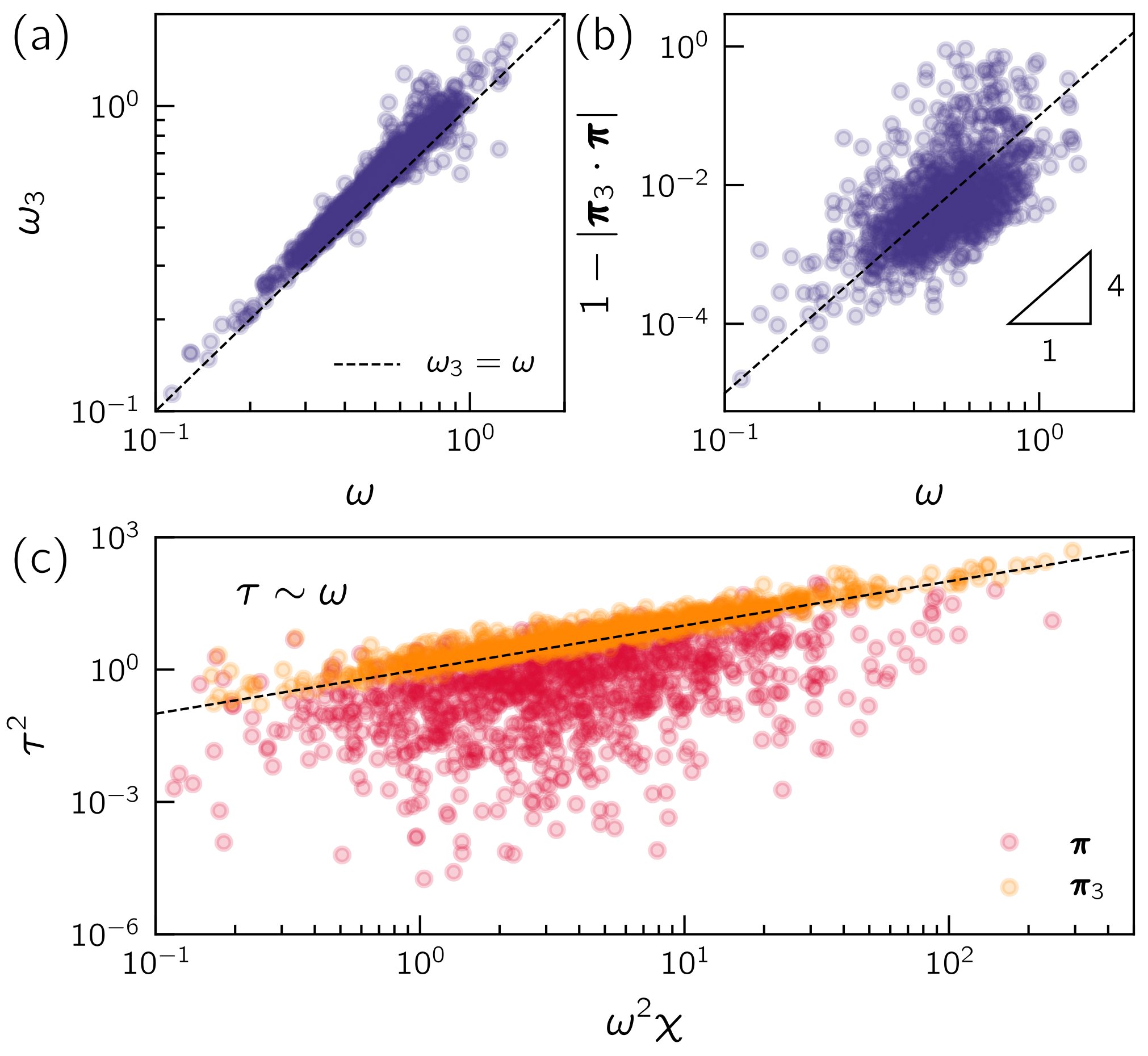}
\caption{\footnotesize (a) Cubic mode frequencies versus PHM frequencies for 1000 different modes in ductile polydisperse glasses ($T_p=0.7$). (b) Convergence of the structure of cubic and pseudo harmonic modes. (c) Relation between the potential energy expansion coefficients. The fourth order coefficient $\chi=\frac{\partial ^4U}{\partial\xv\partial\xv\partial\xv\partial\xv}::\piv\piv\piv\piv$ is found to be independent of the frequency, indicating a linear relation between the asymmetric third-order coefficient $\tau$ and the frequency $\omega$ for cubic modes. This relation does not hold for PHMs. }
\label{fig:apfig3}
\end{figure}

\section{Cubic versus pseudo harmonic modes}
\label{cubevspseudo}

We provide below an energetic and structural comparison between cubic and PHMs. We find that cubic modes are nearly always slightly stiffer than PHMs, see Fig.~\ref{fig:apfig3}a. We expect the structural deviation from cubic and PHMs to scale as $\sim \omega^4$ (the same as obserbed between  PHMs and harmonic modes), which we demonstrate in Fig.~\ref{fig:apfig3}b. As discussed in Ref., cubic modes are the most informative objects from a micromechanics point of view. The reason is that they maximize the asymmetric third-order coefficient $\tau=\frac{\partial ^3U}{\partial\xv\partial\xv\partial\xv}\tripleCdot\piv\piv\piv$, resulting in a better direction to cross the nearby barrier in the energy landscape. As shown Ref.~\cite{episode_1_2020}, cubic modes also display the asymptotic scaling $\tau\sim\omega$ (for $\omega\to0$), which combined with the gapless density of states $D(\omega)\sim\omega^4$ at low frequencies, gives the prediction that the pseudogap exponent $\theta$ of the distribution of thresholds $P(x)\sim x^{\theta}$ is equal to $2/3$, where $x$ is the distance of a mode to the next plastic instability. As shown in Fig.~\ref{fig:apfig3}c, this relation does not hold for PHMs.


\begin{thebibliography}{74}
\bibitem [{foo()}]{footnote}%
  \BibitemOpen
  \href@noop {} {\bibfield  {journal} {\bibinfo  {journal} {In models in which
  pairs of interacting particles are not well-defined, a Voronoi tesselation
  can be used to define pairs. We have checked that this modification does not
  significantly alter the solutions $\piv$ or their energetic or structural
  properties}\ }}\BibitemShut {NoStop}%
\end{thebibliography}
%

\end{document}